\documentclass[pra,twocolumn,preprintnumbers,amsmath,amssymb,floatfix,superscriptaddress,showpacs]{revtex4-1}
\usepackage{graphicx}
\usepackage{graphics}
\usepackage{psfrag}
\usepackage{hyperref}
\usepackage{multirow}
\usepackage[sort&compress]{natbib}
\usepackage{color}
\usepackage{slashed}

\newcommand{\vare}{\varepsilon}

\newcommand{\rmi}{{\rm i}}

\newcommand{\eq}[1]{Eq.~(\ref{#1})}
\newcommand{\fig}[1]{Fig.~\ref{#1}}

\newcommand{\app}[1]{App.~\ref{#1}}

\renewcommand{\sec}[1]{Sec.~\ref{#1}}

\begin{document}

\hypersetup{pdftitle={Phase structure of spin-imbalanced unitary Fermi gases}}
\title{Phase structure of spin-imbalanced unitary Fermi gases}
\date{\today}
\author{I. Boettcher}
\affiliation{Institute for Theoretical Physics, Heidelberg University, D-69120 Heidelberg, Germany}
\author{J. Braun}
\affiliation{Institut f\"{u}r Kernphysik (Theoriezentrum), Technische Universit\"{a}t Darmstadt, D-64289 Darmstadt, Germany}
\affiliation{ExtreMe Matter Institute EMMI, GSI Helmholtzzentrum f\"{u}r Schwerionenforschung mbH, D-64291 Darmstadt, Germany}
\author{T. K. Herbst}
\affiliation{Institute for Theoretical Physics, Heidelberg University, D-69120 Heidelberg, Germany}
\author{J. M. Pawlowski}
\affiliation{Institute for Theoretical Physics, Heidelberg University, D-69120 Heidelberg, Germany}
\affiliation{ExtreMe Matter Institute EMMI, GSI Helmholtzzentrum f\"{u}r Schwerionenforschung mbH, D-64291 Darmstadt, Germany}
\author{D. Roscher}
\affiliation{Institut f\"{u}r Kernphysik (Theoriezentrum), Technische Universit\"{a}t Darmstadt, D-64289 Darmstadt, Germany}
\author{C. Wetterich}
\affiliation{Institute for Theoretical Physics, Heidelberg University, D-69120 Heidelberg, Germany}
\affiliation{ExtreMe Matter Institute EMMI, GSI Helmholtzzentrum f\"{u}r Schwerionenforschung mbH, D-64291 Darmstadt, Germany}

\begin{abstract}
We investigate the phase structure of spin-imbalanced unitary Fermi gases beyond
mean-field theory by means of the Functional Renormalization Group. 
In this approach, quantum and thermal fluctuations are resolved in a 
systematic manner. The discretization of the effective potential on a grid
allows us to accurately account for both first- and second-order phase
transitions that are present on the mean-field level. We compute the 
full phase diagram in the plane of temperature and spin-imbalance and discuss 
the existence of other conjectured phases such as the Sarma phase and a
precondensation  region. In addition, we explain on a qualitative level how we
expect that in-situ density images are affected by our findings and which
experimental signatures may potentially be used to probe the phase structure.
\end{abstract}


\maketitle

\section{Introduction}
The ever-growing advances in the experimental probe and control of ultracold
quantum gases continuously enrich our understanding of strongly-correlated
quantum systems \cite{RevModPhys.80.885}. In particular, the preparation of
locally equilibrated many-body systems allows for the exploration of their
thermodynamic properties such as the phase structure or the equation of state. 
Furthermore, the experimental determination of these key observables 
facilitates the solid benchmarking of theoretical methods for interacting 
quantum systems. 

By populating distinct hyperfine states of a specific class of fermionic
atoms (e.g.~$^6$Li or $^{40}$K), it is possible to emulate two- or
higher-component fermion
systems 
~\cite{2002Sci...298.2179O,*PhysRevLett.92.150402,*Kinast25022005,PhysRevLett.91.020402,Jochim19122003,
*2004PhRvL..92l0401B,*Chin20082004,PhysRevLett.92.040403,PhysRevLett.92.120403,
*2005Natur.435.1047Z}
(see Refs.~\cite{ketterle-review,RevModPhys.80.1215} for reviews). This enables
the realization of ensembles which are reminiscent of a variety of many-body
systems at very  different energy scales such as solid state materials or
neutron stars and gives us an unprecedented opportunity to study the effects
of, e.g., spin-imbalance and temperature in strongly coupled
systems 
~\cite{Zwierlein27012006,*2006Natur.442...54Z,
*PhysRevLett.97.030401,*Schunck11052007,
*2008Natur.451..689S,Partridge27012006, *PhysRevLett.97.190407, Horikoshi22012010,Salomon,Ku03022012}.
The  BCS-BEC crossover of two-component fermions close to an atomic Feshbach
resonance smoothly interpolates between a superfluid of Cooper pairs and a
condensate of composite bosons. In three dimensions, the strongly-coupled
unitary Fermi gas (UFG), where the s-wave scattering length diverges, is
realized at resonance. Here, the scale for all physical observables is set
solely by the Fermi momentum. The high precision in this universal regime on the
experimental side opens up the possibility for detailed benchmarks of the large
variety of available theoretical methods, such as (Quantum) Monte Carlo
calculations~\cite{Carlson:2003zz,*PhysRevLett.93.200404,*PhysRevB.76.165116,
*Bulgac:2005pj,*PhysRevLett.96.160402,*Wingate:2006wy,*PhysRevA.78.023625,*Drut:2008,
*PhysRevA.82.053621},
$\epsilon$-expansions~\cite{PhysRevA.74.053622, *Nishida:2006br, *Nishida:2006eu, *PhysRevA.75.043605,*PhysRevA.75.043620}, 
$T$-matrix approaches~\cite{Haussmann,*PhysRevLett.85.2801,*PhysRevB.61.15370,*PhysRevLett.92.220404,*PhysRevB.70.094508}, 
Dyson--Schwinger equations~\cite{Diehl:2005ae, *PhysRevA.73.033615,
*PhysRevA.77.023626}, $1/N$-expansions~\cite{2007PhRvA..75d3614V},  
2-particle irreducible methods~\cite{Haussmann:2007zz},
renormalization group flow  equations~\cite{Birse:2004ha, PhysRevA.76.021602,
*Diehl:2007ri, *PhysRevA.75.033608, *2008PhRvL.100n0407G,
*Floerchinger:2009pg,*ANDP:ANDP201010458, *Scherer:2010sv,
*2011FBS....51..153F,Bartosch:2009zr,Boettcher201263}, ladder resummation techniques \cite{Kaiser:2011cg,*Kaiser:2013bua}
and exact as well as universal
relations~\cite{PhysRevLett.92.090402,*Tan20082971,*Tan20082952, *Tan20082987}.

In conventional superconductors, a sufficiently large imbalance between spin-up
and spin-down electrons destroys superconductivity due to the mismatch of
the associated Fermi energies. Such a polarization can be realized in a solid
state material by the application of an external magnetic Zeeman field. Since
for ultracold atoms the effective spin degree of freedom originates from  their
individual hyperfine state, this spin-imbalance can be tuned at will by means of
a difference in population. In a microscopic model, this manifests itself in a
difference in chemical potentials. Hereafter, $\mu_1$ and $\mu_2$ denote the
chemical potentials of atoms in  state $|1\rangle$ and $|2\rangle$,
respectively. We assume the former to be the majority species, i.e.
$\mu_1\geq\mu_2$, without loss of generality.

While the ground state of the spin-balanced UFG is commonly believed to be a
homogeneous superfluid, the phase structure in the imbalanced case is less
clear. In fact, given $\mu_1>0$, the density of minority atoms vanishes 
for $\mu_2 \lesssim -0.6\mu_1$~\cite{Chevy,Lobo:2006,Chevy:2006,CRLC,
Bulgac:2007,PS:2007,PG:2007,Schmidt:2011zu}. 
This suggests that superfluidity has to break down at a finite
critical value of the spin-imbalance. For a BCS superfluid, this already
happens for an exponentially small mismatch of Fermi surfaces
\cite{PhysRevLett.9.266, Chandrasekhar}. However, since the UFG has less
pronounced Fermi surfaces, the energy gain from pairing might still compensate
the mismatch and hence be  energetically favorable. We shall discuss below that
within our approximation superfluidity at zero temperature persists down to
$\mu_2\simeq0.09\mu_1$, where it vanishes at a first-order phase transition. 

Besides the breakdown of superfluidity, the existence of exotic phases has been
conjectured for the spin-imbalanced UFG. In the mean-field approximation
\cite{PhysRevLett.96.060401,Gubbels:2006zz,2007NatPh...3..124P}, the homogeneous
Sarma phase \cite{Sarma19631029}, a homogeneous superfluid with 
gapless fermionic excitations, is unstable at zero temperature.
This scenario has been found to persist upon inclusion of bosonic fluctuations~\cite{Boettcher:2014xna}.
Furthermore, inhomogeneous phases such as the Fulde--Ferrell- or 
Larkin--Ovchinnikov-states \cite{PhysRev.135.A550,Larkin:1964zz} may be energetically favored
over the homogeneous superfluid ground state. Hence, such inhomogeneities 
have to be taken into account for a complete study of the 
phase structure. This, however, is beyond the scope of the present 
work and we restrict our discussion to homogeneous phases only.

In this work we study the phase structure of the spin-imbalanced
three-dimensional UFG beyond the mean-field approximation by means of the
Functional Renormalization Group (FRG). This allows us to include the
effect of bosonic fluctuations onto the many-body state.
Besides a large quantitative improvement, such an analysis of fluctuation
effects is also mandatory for a solid understanding of the qualitative features
of the phase diagram. In fact, it is known that the mean-field approximation
fails to predict the correct order of phase transitions in some cases.
In addition, a commonly encountered situation is the suppression of long-range 
order due to fluctuations of the Goldstone modes which can be captured
by our RG approach. Finally, we note that our approach does not suffer from the 
infamous sign problem which complicates {\it ab-initio} Monte-Carlo calculations
of imbalanced systems. To surmount this problem, new techniques have recently
been developed~\cite{Braun:2012ww,Roscher:2013aqa} and successfully applied
to imbalanced Fermi gases~\cite{Braun:2014pka}. From this point of view, our 
present study may also provide useful guidance for future studies of the phase
diagram of spin-imbalanced Fermi gases with Monte-Carlo simulations.

This paper is organized as follows: In \sec{sec:Model} we provide details about
our studied system and its phase structure on the mean-field level. Next, we
discuss the truncation and numerical implementation of the FRG setup used to
include fluctuations beyond the mean-field level. Results on the phase structure
of the imbalanced UFG including fluctuations are presented in 
\sec{sec:Results}. We discuss experimental signatures reflecting the
phase diagram in \sec{SecExpSig}. Our concluding remarks are given in 
\sec{sec:Conclusion}.

\section{Model}
\label{sec:Model}
We consider two-component ultracold fermions close to a broad s-wave Feshbach
resonance (FR).  The scattering physics can be described by the two-channel 
model \cite{gurarie-review,Diehl:2005ae,PhysRevA.73.033615}, where the closed 
channel is incorporated by means of a bosonic field $\phi$. For a broad FR this 
is equivalent to a purely fermionic one-channel model: both result in the same
universal low-energy physics. In the purely fermionic picture, the bosons emerge
as a pairing (or order-parameter) field in the particle-particle
channel, $\phi \sim \psi_1\psi_2$.
The assumption of a broad FR is valid, e.g., for $^6$Li, where the
resonance is located at $B_0=832.2$\,G with a width $\Delta B\simeq
200$\,G \cite{PhysRevLett.110.135301}.

The microscopic action of the two-channel model reads
\begin{align}
 \nonumber &S[\psi_\sigma,\phi] = \int_X\Bigl[\sum_{\sigma=1,2}\psi^*_
  \sigma(\partial_\tau-\nabla^2-\mu_\sigma)\psi_\sigma \\
 \label{Mod1} &\mbox{ }+\phi^*(\partial_\tau -\nabla^2/2 +\nu_\Lambda)\phi - 
g(\phi^*\psi_1\psi_2+\text{h.c.})\Bigr].
\end{align}
It serves as the starting point for our computations. The atoms in hyperfine 
state $|\sigma\rangle$ are represented by a Grassmann-valued field
$\psi_\sigma(\tau,\vec{x})$ with imaginary time $\tau$
\cite{negele-book,altland-book}. We employ units such that $\hbar=k_{\rm
B}=2M=1$, where $M$ is the mass of the atoms.  The imaginary time domain is
compactified to a torus of circumference $T^{-1}$ in the standard way and we
write $\int_X = \int\mbox{d}\tau\int\mbox{d}^3x$. 

We allow for an imbalance in the chemical potentials of the individual species,
$\mu_1$ and $\mu_2$, respectively. Moreover, we assume the 1-atoms to be the
majority species such that
\begin{align}
\label{Mod2} \delta\mu = h = \frac{\mu_1-\mu_2}{2} \geq0\,,\ 
\mu=\frac{\mu_1+\mu_2}{2}\,.
\end{align}
The spin-imbalance, $\delta\mu$, is frequently also referred to as Zeeman field,
$h$. We can thus write $\mu_{1,2}=\mu\pm\delta\mu$, where $\mu$ is the average 
chemical potential in the system. 

The model in \eq{Mod1} is valid on momentum scales much smaller than a 
(large) momentum cutoff $\Lambda$. In practice one can choose
$\Lambda$ to be sufficiently large compared to the many-body scales determined
by density or temperature, but well below the inverse van-der-Waals length.
Details of the interatomic interaction are then irrelevant. We further assume
the interactions to be of zero range.  The detuning from resonance,
$\nu_\Lambda\propto(B-B_0)$, has to be fine-tuned such that $a^{-1}=0$. 
With this renormalization, thermodynamic observables become independent of 
$\Lambda$ and $a^{-1}$. The Feshbach coupling $g^2\propto \Delta B$ is related
to the width of the resonance, which we assume to be large in the following.

We employ a functional integral representation of the quantum effective 
action, $\Gamma[\psi_\sigma, \phi]$, in terms of coherent states. The effective
action is the generating functional of one-particle irreducible correlation
functions. When evaluated at its minimal configuration, it is related to the
partition function according to $\Gamma_0=-\ln Z(\mu,\delta\mu,T)$. For a
comprehensive introduction to functional methods in the context of ultracold
atoms see, {e.g., Refs.}~\cite{Stoof-book,Boettcher201263}.

In the present approach, the fermion fields only appear quadratically and can be
integrated out, leaving us with a description in terms of the pairing field,
$\phi$, only. In the BCS-BEC crossover, the pairing field has the intuitive
interpretation of Cooper pairs or composite diatomic molecules in the BCS- and
BEC-limits, respectively. For the UFG, however, such a simple picture has not
been found yet. Loosely speaking, the many-body state in this limit rather is a
strongly-correlated quantum soup with both bosonic and fermionic features. We
assume the boson field expectation value to be homogeneous in the following,
$\phi_0\neq \phi_0(\vec{x})$. Below we discuss why this should be a reasonable
assumption for the spin-imbalanced UFG. 
A non-vanishing field expectation value $\phi_0(\mu,\delta\mu,T) = 
\langle \phi \rangle \neq 0$ then signals superfluidity of the system.

The field expectation value is determined by the minimum of the full effective 
potential, $$U(\rho=\phi^*\phi) = \Omega^{-1}\, \Gamma[\phi]\,,$$ where $\Omega=V/T$
with the three-dimensional volume $V$ and $\phi=\,$const. Due to global
U(1)-invariance of the microscopic action, the effective potential depends
only on the U(1)-invariant $\rho=\phi^*\phi$. Without loss of generality we
assume $\phi_0$ to be real-valued. For fixed $\mu$ and $T$, the amplitude of
$\phi_0$ is a (not necessarily strictly) monotonously decreasing function of
$\delta\mu$. At the critical imbalance $\delta\mu_{\rm c}(\mu,\,T)$\,, the 
global minimum of $U(\rho)$ approaches $\rho_0=0$ either discontinuously or
continuously, resulting in a first- or second-order phase transition,
respectively.

\begin{figure}
  \includegraphics[width=.45\textwidth]{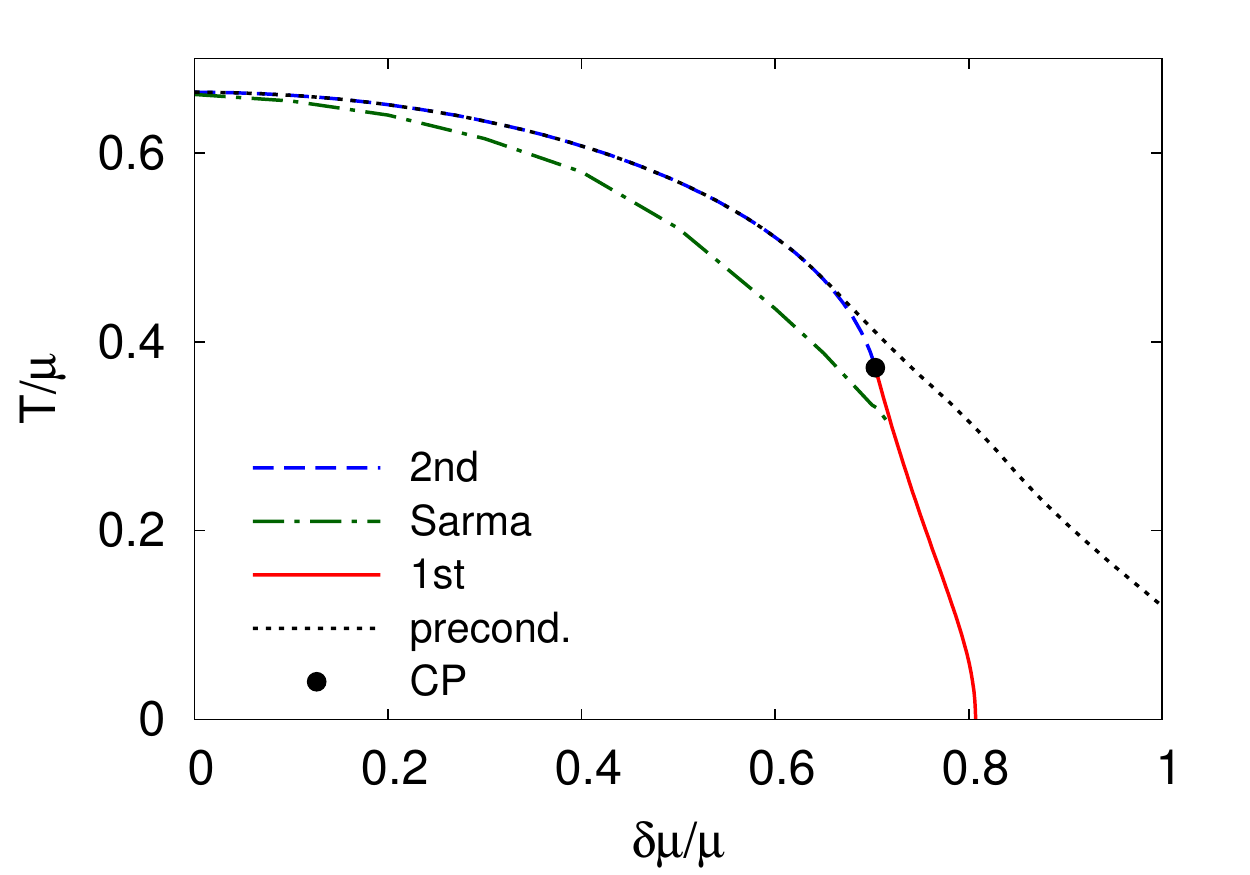}
  \caption{(Color online) Mean-field phase diagram of the spin-imbalanced UFG.
    We show the superfluid-to-normal transition and the Sarma crossover.
    The location of the phase boundaries can be obtained from the
    mean-field expression for the grand canonical potential, \eq{Mod3}, or from 
    the FRG flow by omitting bosonic fluctuations. In the 
    latter approach it is possible to additionally resolve the precondensation 
    line (black, dotted line), below which a minimum at $\Delta_{0,k}$
    appears during the RG flow, but vanishes for $k\to 0$, thereby leaving the 
    system in the normal phase. We discuss the phenomenon of precondensation in 
    \sec{sec:Results} in detail.}
  \label{fig:phaseMFA}
\end{figure}

In the mean-field approximation the effective action is computed from a 
saddle-point approximation of the functional integral, here under the assumption
of a homogeneous field expectation value. For convenience, we parametrize
it in terms of the gap parameter, $\Delta^2 =g^2\rho$, rather than $\rho$
itself. For the UFG we find
\begin{align}
 \label{Mod3} &U(\Delta^2,\mu,\delta\mu,T) = \int \frac{\mbox{d}^3q}{(2\pi)^3} 
\Biggl[ \Bigl(q^2-\mu-E_q-\frac{\Delta^2}{2q^2}\Bigr)\\
 \nonumber &\mbox{ }-T \ln\Bigl(1+e^{-(E_q-\delta\mu)/T}\Bigr)-T \ln\Bigl(1
 +e^{-(E_q+\delta\mu)/T}\Bigr)\Biggr]\,,
\end{align}
with $E_q=\sqrt{(q^2-\mu)^2+\Delta^2}$, cf. e.g.
\cite{PhysRevLett.96.060401,2007NatPh...3..124P,PhysRevA.82.013624}. The 
mean-field phase boundary is found from the global minimum, $\Delta_0^2$, of the
effective potential $U(\Delta^2)$.
Note that the mean-field approximation can be recovered from the 
FRG equation when all bosonic fluctuations are neglected, see our discussion 
below. This allows us to study the impact of fluctuations in a unified 
approach. We show the mean-field phase diagram in Fig.~\ref{fig:phaseMFA}.

The mean-field analysis predicts a first order phase transition at zero 
temperature at $\delta\mu_{\rm c}/\mu = 0.807$ (red, solid line). 
This is often referred to as Chandrasekhar--Clogston limit 
\cite{Chandrasekhar, PhysRevLett.9.266}.
At this point, the field expectation value jumps from $\Delta_0/\mu=1.162$ to
zero. The second order phase transition (blue, dashed line) of the balanced case
occurs at $T_{\rm c}/\mu=0.665$. As a reaction to non-zero spin-imbalance, the
transition changes from second to first order at the tricritical point
$(\delta\mu_{\rm CP}/\mu,\, T_{\rm CP}/\mu)~=~(0.704,\,
0.373)$ in agreement with previous findings, see Refs.~\cite{Chevy:2010zz,
Gubbels:2013mda} for reviews.
Also shown in \fig{fig:phaseMFA} are the Sarma crossover (green,
dot-dashed line) and so-called precondensation line (black, dotted line). These
features are discussed in detail in \sec{sec:Results} when we compare the
mean-field phase structure to the results from our RG analysis including
fluctuations.

\section{Functional Renormalization Group}
\label{sec:FRG}

We now include the feedback of bosonic fluctuations onto the effective potential.
This is particularly important for the regime with spontaneously broken
symmetry, where a massless Goldstone mode appears. The FRG approach allows to
systematically include the effect of the latter and is free of infrared
divergences, see e.g. Refs. \cite{Litim:1998nf, *Berges:2000ew,*Gies:2006wv,
*Schaefer:2006sr, *Pawlowski20072831, *Delamotte:2007pf, *Kopietz2010,
*Metzner:2011cw, *Braun:2011pp,*vonSmekal:2012vx} for a general introduction to
the method, and Refs. \cite{2008arXiv0801.0009B, PhysRevA.76.021602,
*Diehl:2007ri,*PhysRevA.75.033608,*2008PhRvL.100n0407G,*Floerchinger:2009pg,
*ANDP:ANDP201010458,*Scherer:2010sv,*2011FBS....51..153F,Bartosch:2009zr,
Boettcher201263} for an overview on the application in the cold atoms context.

The FRG is based on an exact flow equation for the effective average action
$\Gamma_k[\psi_\sigma,\phi]$. The latter interpolates smoothly between the
microscopic action at large momentum scales and the full quantum effective
action at low momentum scales, $\Gamma_{k=\Lambda}=S$ and $\Gamma_{k=0}=\Gamma$,
respectively. Herein $k$ is a flowing momentum scale, and $S$ is given by the
microscopic model in \eq{Mod1} for the present analysis. The flow
equation reads
\begin{align}
  \label{frg1} \partial_k \Gamma_k[\psi_\sigma,\phi] = \frac{1}{2}\mbox{STr} 
  \Bigl( \frac{1}{\Gamma_k^{(2)}[\psi_\sigma,\phi]+R_k}\partial_k R_k\Bigr),
\end{align}
where $\Gamma_k^{(2)}$ is the second functional derivative with respect to the 
field content of the theory, and $R_k$ is an infrared regulator
\cite{Wetterich1993}. Accordingly,
$(\Gamma_k^{(2)}+R_k)^{-1}$ is the full propagator of the regularized theory.
$\mbox{STr}$ denotes a supertrace, see e.g. the detailed discussion 
in~\cite{PhysRevA.89.053630}. 
\eq{frg1} is an exact equation that is very convenient for practical
purposes due to its one-loop structure. However, the presence of the full
propagator on the right-hand side makes the use of truncations necessary in
most cases of interest. 

The effective average action is a functional of the mean-fields, $\psi_\sigma$ 
and $\phi$, of fermions and bosons, respectively. To
approximately resolve its functional form we apply the ansatz
\begin{align}
 \nonumber \Gamma_k[\psi_\sigma,\phi] &=\int_X \Bigl( \psi_\sigma^* P_{\psi
  \sigma, k}(\partial_\tau,-\rmi \nabla) \psi_\sigma \\
 \nonumber &\mbox{} + \phi^* P_{\phi,k}(\partial_\tau,-\rmi \nabla)
    \phi + U_k(\rho=\phi^*\phi)\\
 \label{truncation} &-\mbox{ }g_k (\phi^*\psi_1\psi_2+\text{h.c.})\Bigr).
\end{align}
In this way we parametrize the system in terms of the inverse fermion and boson
propagators ($P_{\psi\sigma}$ and $P_\phi$), the effective potential
($U$), and the Feshbach coupling ($g$). The scheme used in this work builds on a
scale-dependent derivative expansion of the boson propagator, while keeping the
fermion propagator in its microscopic form. Accordingly, we have
\begin{align}
 P_{\psi \sigma, k}(Q) &= P_{\psi\sigma, k}(\rmi q_0,\vec{q}) = \rmi q_0
+ q^2 -\mu_\sigma\,,\\
 P_{\phi, k}(Q) &= P_{\phi, k}(\rmi q_0,\vec{q}) = A_{\phi,k} \Bigl( \rmi
q_0 + \frac{q^2}{2}\Bigr)\,.
\label{eq:Propagators}
\end{align}
Systematic extensions of this truncation are possible and yield 
quantitative improvement, see our discussion below and, e.g., 
\cite{PhysRevA.89.053630}.

The key ingredient of our analysis is keeping the full functional form of the 
effective average potential $U_k(\rho)$. In this way we are able to properly 
resolve first-order phase transitions and also to quantitatively improve 
results beyond a Taylor expansion of $U_k(\rho)$ in powers of the field. The 
flow equation for the effective potential is obtained from \eq{frg1} for 
a constant background field $\phi$. It is given by
\begin{align}
\label{frg2} \dot{U}_k(\rho) = \dot{U}_k^{(F)}(\rho)
  + \dot{U}_k^{(B)}(\rho),
\end{align}
where the superscripts $F$ and $B$ indicate the contributions from fermionic 
and bosonic loops, respectively. The dot denotes a derivative with respect to
RG-time $t=\ln(k/\Lambda)$. 
We discuss \eq{frg2} in detail in \app{app:Flow}. Here we focus on the most 
important aspect for the present analysis, which is the interplay between 
$\dot{U}^{(F)}$ and $\dot{U}^{(B)}$.

The explicit form of the beta function on the right-hand side of \eq{frg2}
depends on the choice of truncation for $\Gamma_k$, and the regulators $R_\phi$
and $R_{\psi\sigma}$ for bosons and fermions, respectively. To exemplify
the key features we now discuss its form obtained for the truncation 
(\ref{truncation}) with the regulators from Eqs.~(\ref{frg7}) and (\ref{frg9}). 
The general equation is displayed in App.~\ref{app:Flow}. 
For clarity, we restrict the formulas to the zero temperature case 
for the moment. We then find 
\begin{align}
 \label{frg3}\dot{U}_k^{(F)}(\rho) &= -\frac{k^{2}}{3\pi^2\sqrt{1+g^2\rho/k^4}} 
    \theta\Bigl(\sqrt{k^4+g^2\rho}-\delta\mu\Bigr)\\
 \nonumber &\times\Bigl[(\mu+k^2)^{3/2}\theta(\mu+k^2)-(\mu-k^2)^{3/2}\theta(\mu-k^2)\Bigr]
\end{align}
for the fermionic part, where $\theta(x)$ is the Heaviside step function. There 
is no direct feedback of $U_k(\rho)$ onto its flow from the expression in
\eq{frg3}. In fact, the integration of only this contribution yields the
mean-field effective potential.
However, including also bosonic fluctuations has indirect impact on 
$\dot{U}^{(F)}$ 
owing to the running Feshbach coupling $g_k$.

Due to the presence of the bosonic contribution in \eq{frg2}, we are faced with a 
coupled flow equation where fermionic and bosonic terms compete. 
The corresponding flow at zero temperature reads
\begin{align}
  \label{frg4}\dot{U}_k^{(B)}(\rho) &= \frac{\sqrt{2}\, k^{5}}{3\pi^2} 
\Bigl(1-\frac{\eta_\phi}{5}\Bigr)\\
 \nonumber &\mbox{ }\times \frac{A_\phi k^2+U_k'(\rho)+\rho
U''_k(\rho)}{\sqrt{(A_\phi k^2+U_k'(\rho))(A_\phi k^2+U_k'(\rho)+2\rho U''_k(\rho))}}\,,
\end{align}
where primes denote derivatives with respect to $\rho$. The appearance of 
$U_k'(\rho)$ and $U_k''(\rho)$ on the right-hand side of the flow equation 
necessitates a good resolution of $U_k(\rho)$ during the flow. For this
purpose we discretize the function $U_k(\rho)$ on a grid of typically
$\geq100$ points.

The initial condition for the flow of $U_k(\rho)$ is given by
$U_\Lambda(\rho)=\nu_\Lambda \rho$\,. During the flow, the effective average
potential acquires a more complex form, which is accounted for by the
discretization on the grid. We keep track of the scale-dependent minimum 
$\rho_{0,k}$, and determine the phase structure from the order parameter
$\Delta_0^2 = \Delta_{0,k=0}^2$ at $k=0$. 

The boson dynamics are encoded in the inverse boson propagator 
$P_{\phi}$, cf. \eq{eq:Propagators}.
Here we apply a scale-dependent derivative expansion, where
$P_{\phi}(Q)$ is expanded in powers of $\rmi q_0$
and $q^2$ for each scale $k$ separately. Due to the presence of the regulator 
in the flow equation, this is expected to give a good approximation of the one-loop 
integral in \eq{frg1}. We will now argue why the simple form
(\ref{eq:Propagators}) is expected to be sufficient to describe the phase 
structure. In general, the leading order terms in the 
expansion of $P_\phi$ read
\begin{align}
 \label{frg5} P_{\phi ,k}(Q) = A_{\phi,k}\Bigl( \rmi Z_{\phi,k}   q_0 + 
  \frac{1}{2}q^2 + V_{\phi, k} q_0^2 +\dots  \Bigr).
\end{align}
At the microscopic scale, $A_{\phi, \Lambda}=Z_{\phi, \Lambda}=1$ and  $V_{\phi,
\Lambda}=0$. By taking two functional derivatives of \eq{frg1} one can
derive the flow equation for $P_\phi$, and thus for the running couplings
parametrizing it, see e.g. \cite{PhysRevA.89.053630}. 

From studies of bosonic systems it is known that $Z_{\phi,k}$ vanishes like
$k^{3-d}$ as $k\to 0$ in $d<3$ spatial dimensions, and  vanishes logarithmically
for $d=3$ \cite{0295-5075-80-5-50007, PhysRevB.77.064504, PhysRevB.78.014522,
PhysRevB.89.035126}.
Hence, the linear frequency term  is replaced by a quadratic frequency
dependence with $V_{\phi,k}>0$ in the infrared Goldstone regime
\cite{PhysRevB.77.064504}. In order to describe the boson dynamics
consistently, both $Z_{\phi, k}$ and $V_{\phi, k}$ need to be taken into
account. Without $V_{\phi, k}$ the propagator at a low scale $k$ would become
frequency-independent. However, in three dimensions the running of $Z_{\phi,
k}$ with $k$ is only logarithmic, and there is no strict need to incorporate 
$V_{\phi,k }$. For instance, at the scale $k$ where
$\Delta_{0,k}$ saturates, $Z_{\phi, k}$ typically still has a substantial size
$\simeq 0.5$, and $V_{\phi,k} q_0^2$ represents a subleading term. Moreover, it
has been demonstrated previously that the inclusion of $Z_{\phi, k}$ only leads
to corrections of a few percent in, e.g., the critical temperature, cf.
\cite{PhysRevA.89.053630}. Furthermore, these modifications are counter-balanced
to some extent by the running of $V_{\phi, k}$. Hence we choose the following
consistent approximation 
\begin{align}
 \label{frg6} Z_{\phi, k} = 1,\ V_{\phi, k} =0
\end{align}
for all $k$. The flow of $A_{\phi, k}$, on the other hand, is
incorporated by means of the anomalous dimension $\eta_{\phi, k}~=~-~\partial_t
\ln A_{\phi, k}$. The corresponding flow equation is given in App. \ref{app:Flow}.

The regulator functions $R_\phi$ and $R_{\psi \sigma}$ which enter the flow
equation~(\ref{frg1}) have to meet several conditions \cite{Berges:2000ew}, but
can otherwise be chosen freely. If there was no truncation of the effective 
average action, the fact that the regulator functions vanish for $k\to 0$
would entail that the result is independent of the regulator. In
practice, the truncation introduces a spurious dependence on the choice of
regulator, which may be employed for an error estimate by comparing results
obtained for different regulators, see e.g. \cite{Schnoerr:2013bk}.

In order to regularize the fermion propagator we apply two regulator choices 
separately. They read
\begin{align}
  \label{frg7}  R_{\psi \sigma}^Q \equiv R_\psi^Q &=( k^2\mbox{sgn}(\xi_q) -
    \xi_q)\theta(k^2-|\xi_q|)
\end{align}
and
\begin{align}
 \label{frg8} R_{\psi \sigma}^Q &=( k^2\mbox{sgn}(\xi_{q\sigma}) -
    \xi_{q\sigma})\theta(k^2-|\xi_{q\sigma}|)
\end{align}
with $\xi_q=q^2-\mu$ and $\xi_{q\sigma}=q^2-\mu_\sigma$, respectively. Both 
forms regularize only spatial momenta, $q^2$, and constitute a generalization of 
the fermion regulators used for the balanced case in previous works. Remarkably, 
the choice (\ref{frg7}), where both species are regularized around the average
chemical potential $\mu$, is sufficient to render all flows finite
and furthermore allows to derive analytic flow equations for both $U_k(\rho)$
and $A_{\phi, k}$. In contrast, for the second choice, \eq{frg8}, the loop
integral has to be performed numerically. We find that the resulting
phase diagrams for both choices coincide within the numerical error, see 
\app{app:Phase}.

For the bosons we use
\begin{equation}
 \label{frg9} \bar{R}_\phi^Q = A_\phi R_\phi^Q = A_\phi (k^2-q^2/2) \theta(k^2-q^2/2)\,,
\end{equation}
as in previous balanced case studies. The full set of flow equations for the
running couplings is given in \app{app:Flow}.

We restrict this investigation to the stability of homogeneous superfluid order.
A competing effect from inhomogeneous order is expected to show precursors in
the renormalization group flow.
One of those is the vanishing of $A_{\phi,k}$ at some non-zero momentum
scale $k>0$ \cite{PhysRevB.80.014436}. At this point, the truncation employed 
here would become insufficient. Since we do not detect signs of 
such a behavior anywhere near the superfluid phase, it seems reasonable to 
restrict ourselves to a homogeneous order parameter $\Delta_0 \neq 
\Delta_0(\vec{x})$.  A more detailed discussion of the
appearance of inhomogeneous order in the presence of spin- and mass-imbalance
from an FRG perspective will be given in \cite{Braun:2014}.

\section{Results}
\label{sec:Results}

\begin{figure}
  \includegraphics[width=.45\textwidth]{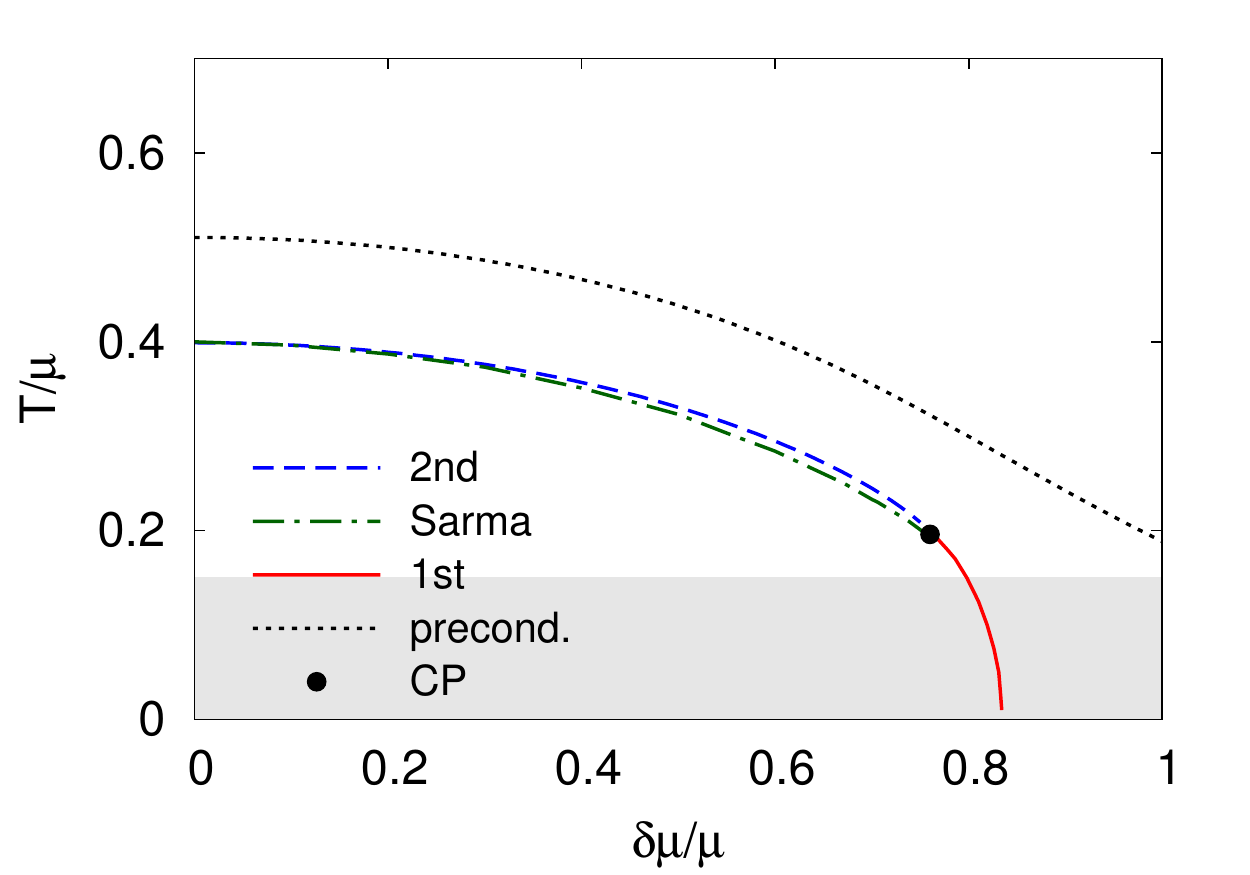}
  \caption{(Color online) Phase diagram of the spin-imbalanced UFG beyond the mean-field
    approximation. The phase boundaries are obtained from the FRG evolution of
    the effective potential including the feedback of bosonic fluctuations. The
    critical temperature of the balanced system is found to be $T_{\rm
    c}/\mu=0.40$. 
    For small $\delta\mu/\mu$ we find a second order phase transition with a
    reduced critical temperature.
    For low temperatures, spin-imbalance results in a breakdown
    of superfluidity by means of a first-order phase transition. We extract
    $\delta\mu_{\rm c}/\mu=0.83$ for the critical imbalance at zero temperature.
    The second-order line terminates in a tricritical point (CP). 
    We indicate the Sarma crossover by the green, dash-dotted line. The region
    between the precondensation line (black, dotted) and the phase boundary
    gives an estimate for the pseudogap region as is explained in the main
    text.}
  \label{fig:phaseFRG}
\end{figure}

We now discuss the phase structure of the system as obtained from the FRG setup
described in the previous section.
In particular, we show results computed with the fermionic regulator
\eq{frg7} from above. As we demonstrate in \app{app:Phase} below, these results
agree very well with the ones obtained using the regulators \eq{frg8}, but are
numerically more stable due to the analytic expressions for the beta
functions.

\subsection{Phase diagram}

The  phase diagram of the spin-imbalanced UFG beyond mean-field theory is shown
in \fig{fig:phaseFRG}. The overall phase structure is qualitatively similar to
the mean-field result, cf. \fig{fig:phaseMFA}. Noticeably, however, the critical
temperature is reduced drastically when fluctuations are included. In
the balanced limit we find a second-order phase transition (blue, dashed
line) with $T_c/\mu=0.40$\,. This is in good agreement with recent measurements
\cite{Ku03022012} as well as QMC calculations \cite{Wingate:2006wy} 
and consistent with previous FRG calculations based on a 
Taylor expansion of the effective potential \cite{PhysRevA.89.053630}. 

As the spin-imbalance is increased, the transition changes from second to first
order in  a tricritical point located at $(\delta\mu_{\rm CP}/\mu, T_{\rm
CP}/\mu) = (0.76, 0.20)\,.$
Below this point we find a first-order transition line, which appears to extend
down to $T\approx0$ (red, solid line). From an extrapolation of the
transition line computed for $T\geq 0.01$, we deduce a first-order phase
transition for $\delta\mu_{\rm c}/\mu=0.83$ at vanishing temperature.
This is in reasonable agreement with the recent experimental finding of a
first-order transition at $\delta\mu_{\rm c}/\mu=0.89$
\cite{PhysRevA.88.063614}.

Notably, the critical imbalance at zero temperature lies above the 
mean-field value
($\delta\mu_c^{\rm MFA}/\mu=0.807$). This is an interesting observation since
usually bosonic fluctuations tend to destroy ordering. In the present case, however, the non-trivial feedback of those fluctuations into the flow also modifies the fermionic ``mean-field''
contributions from $\dot{U}_k^{(F)}$. In this way, for large enough $\delta\mu$ and small enough $T$,
the non-trivial minimum of $U_k$ is stabilized rather than washed out. This illustrates how the competition
of fermionic and bosonic contributions results in non-trivial effects on the phase structure of the system.
Unfortunately, the nonlinear structure of the FRG flow equations inhibits a straightforward interpretation
of these observations in terms of customary many-body phenomenology. A more detailed investigation
in this respect is left for future work.


Note that it is numerically impossible to calculate observables at 
exactly $k=0$\,. However, the flow usually freezes out at a finite scale below 
the relevant many-body scales  present in the theory. In order to reliably 
extract the phase structure we may hence stop the integration of the flow 
equation at any sufficiently small $k$ such that $\Delta_{0,k}\simeq
\Delta_{0,k=0}$ is frozen out. Especially in the first-order region at
 low temperatures $T/\mu\lesssim 0.15$, the complexity of the flow 
equation makes it harder to reach the deep infrared. Due to accumulating
numerical errors, the flow needs to be stopped at relatively high 
$k<1$\,. This entails that a sufficient convergence of $\Delta_{0,k}$ inside the 
superfluid phase might not be achieved yet. However,  
we will argue in \sec{subsec:ScaleEvol} below that the
position of the first-order phase transition is not affected and can still be
determined accurately. A conservative
estimate of the domain where the IR scale is modified is indicated by the gray
band in \fig{fig:phaseFRG}.

Concerning the regulator dependence, we find the result 
for the phase boundary to differ by less than $5\%$ for the
two choices of fermion regulators in Eqs.~(\ref{frg7}) and (\ref{frg8}). We
compare both phase diagrams in \fig{fig:phase_FRG_cmp} and provide a more
detailed discussion in \app{app:Phase}. The insensitivity of the critical
line to the regularization scheme indicates the stability of our predictions
within the given truncation scheme for the effective average action. We would
also like to note here that, at least for small spin imbalances, one may employ
a Taylor expansion for the effective potential $U(\rho)$ as recently done in
Ref.~\cite{Krippa:2014kra}. At least for an expansion up to order $\rho^2$ we
observe that the results for the critical temperature from a Taylor expansion of
the effective potential are larger than those from our study with a discretized
effective potential, which naturally includes higher-order couplings. Moreover,
the difference between the critical temperatures increases with increasing
spin-imbalance, see \fig{fig:phaseTaylor} in \app{app:Phase} for a more detailed
discussion.

In addition to the superfluid-to-normal transition we also show the crossover
to the so-called Sarma phase \cite{Sarma19631029}, which we determine from the
criterion $0<\Delta_0 \leq\delta\mu$. If the latter condition is fulfilled, the
lower branch of the dispersion of fermionic quasiparticles,
\begin{align}
 E_p =\sqrt{(p^2-\mu)^2+\Delta_0^2}-\delta\mu,
\end{align}
extends below zero. Strictly speaking, the Sarma phase is well defined only for
$T=0$ where one finds a momentum interval $[p_{\rm min},p_{\rm max}]$
which is occupied macroscopically. In turn, this also results in gapless
fermionic excitations in the homogeneous superfluid. For an extended
discussion of the Sarma phase in the BCS-BEC crossover we refer to
\cite{Gubbels:2013mda, Boettcher:2014xna}.

Already on the mean-field level the Sarma phase is found to be absent at low
$T$. The Sarma crossover meets the first order transition line just below the
critical point. For lower temperatures, the Sarma criterion cannot be
fulfilled anymore since the gap jumps to zero from $\Delta_0~>~\delta\mu_c$.
This situation persists beyond the mean-field level, as can be seen in
\fig{fig:phaseFRG}.
In fact, the Sarma phase shrinks at low imbalance, occurring only in
the close vicinity of the superfluid-to-normal transition. 
Interestingly, the opposite effect has been observed on the BCS 
side of the crossover in two spatial dimensions: there it is found that the 
inclusion of bosonic fluctuations beyond mean-field theory changes the 
transition from first to second order, entailing the presence of a Sarma phase 
even at $T=0$, see Ref.~\cite{PhysRevX.4.021012}.

\subsection{Scale evolution and precondensation}
\label{subsec:ScaleEvol}

\begin{figure}
  \includegraphics[width=.5\textwidth]{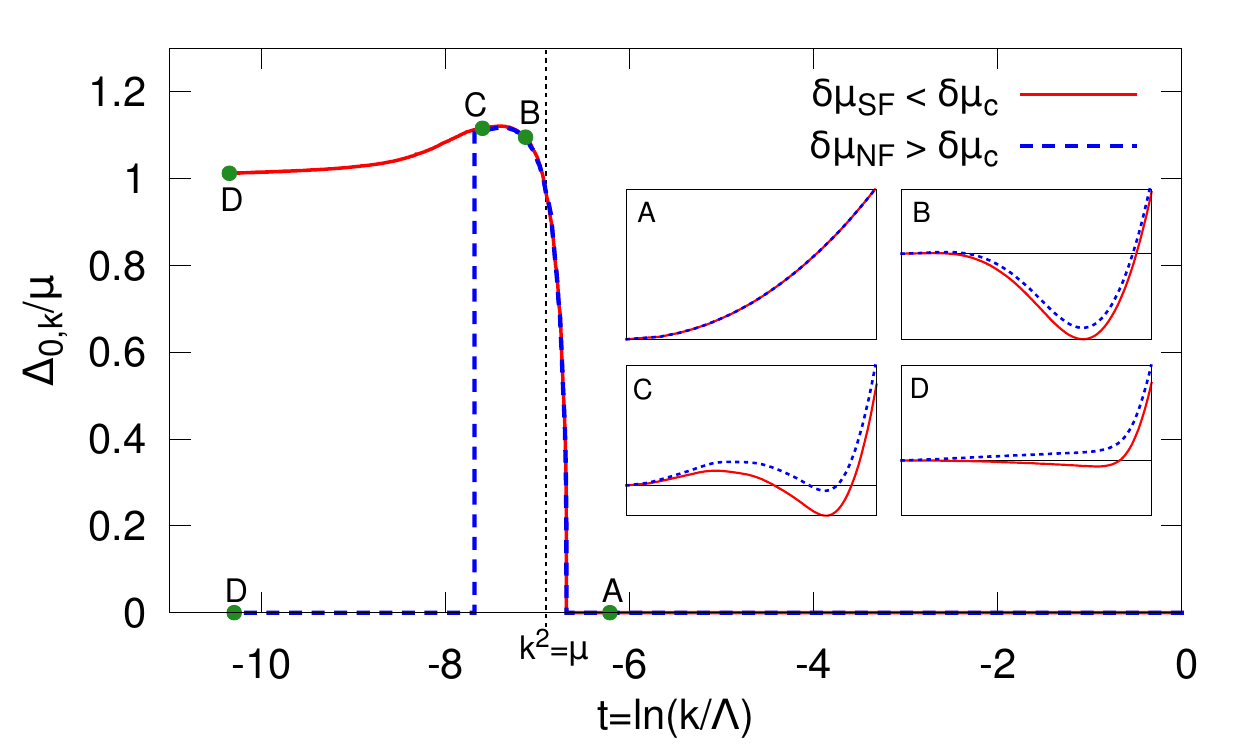}
  \caption{(Color online) Scale evolution of the minimum $\Delta_{0,k}$ of the effective
    average potential close to a first-order phase transition at $\delta\mu_c$. 
    The evolution proceeds from the ultraviolet ($k=\Lambda$, $t=0$) to
    the infrared  ($k\to0$, $t\to-\infty$).
    The insets show the shape of the effective potential $U_k(\Delta)$ at
    several points along the scale evolution. The solid (red) lines correspond
    to a point in the broken phase ($\delta\mu_{\rm{SF}} = 0.78\mu$), where the
    global minimum of the effective potential is non-zero in the
    infrared. The dotted (blue) line represents a point with
    $\delta\mu_{\rm{NF}} = 0.79\mu$, where the global minimum in the
    infrared is located at $\Delta_{0,k=0}=0$. For all plots, $T/\mu = 0.17$.}
  \label{fig:Deltak}
\end{figure}

In \fig{fig:Deltak} we show the scale evolution of the minimum of 
the effective average potential as a function of the RG-scale $k$ for 
fixed $T/\mu = 0.17$ and two different spin-imbalances, $\delta\mu_{\rm{SF}} = 
0.78\mu$ and $\delta\mu_{\rm{NF}} = 0.79\mu$\,.
For large $k$ the running of couplings is attracted to an ultraviolet fixed
point. This scaling regime is left when $k$ becomes of the order of the
many-body scales, see \app{app:Flow}.
For low enough temperatures local symmetry breaking occurs at $k^2\simeq\mu$,
associated with a non-zero minimum of the effective potential,
$\Delta_{0,k}>0\,.$
Competing bosonic and fermionic fluctuations then determine whether the 
non-vanishing gap remains (red, solid line) or vanishes (blue, dotted line)
in the infrared (IR), i.e. for $k\to 0$.

The two values of $\delta\mu$ shown in \fig{fig:Deltak} are chosen such that
they lie on the opposite sides of the first-order phase boundary.
In both cases a non-vanishing gap,  $\Delta_{0,k}~>~0\,,$ is  generated during
the flow at $t_{\rm sb}$.
Only for $\delta\mu=\delta\mu_{\rm{SF}}$ it persists for $t\to -\infty$, leading
to superfluidity (SF) and symmetry breaking in the IR. 
For $\delta\mu=\delta\mu_{\rm{NF}}$ instead, $\Delta_{0,k}$ jumps back to zero
at the finite scale $t_{\rm sr}=-7.69$, below which the symmetry remains
restored such that one finds a normal fluid (NF). In both cases, the effective
potential at intermediate $k$ exhibits two local minima (inset C), but for
$\delta\mu_{\rm{NF}}$ the non-trivial one is raised above $U_k(\rho=0)$
and disappears (inset D) during the flow.

The appearance of a non-zero $\Delta_{0,k}$ in a limited range
$k_{\rm{sr}} < k < k_{\rm{sb}}$ is called {\it precondensation}, see e.g. 
\cite{Boettcher201263}.
It can be interpreted as the formation of pairs and local phase coherence,
although long-range order is destroyed due to fluctuations. The associated
coherence length can be estimated by $k_{\rm{sr}}^{-1}$. 
An analogous phenomenon
exists in relativistic theories \cite{Braun:2009si}.
In \fig{fig:phaseFRG} the precondensation region is enclosed by the black,
dotted line and the phase boundary.

The phenomenon of precondensation is closely
related to pseudogap physics, which refers to a situation where the gas is
in the normal phase, although low-lying fermionic excitations are gapped. We refer to \cite{PhysRevLett.71.3202,
Randeria19981754} for a discussion  of the latter in the context of
superconductivity, and \cite{2010NatPh...6..569G,PhysRevLett.106.060402,PhysRevA.83.053623,2014arXiv1409.4743S}
for pseudogap physics in ultracold Fermi gases. In our case,
superfluidity is absent due to $\Delta_{0}=0$, but excitations with momentum $k_{\rm{sr}} < k < k_{\rm{sb}}$, which
is typically on the order of $k_{\rm F}$, are energetically disfavoured. This leads to a strong suppression in the density of states and thus
of the contribution of these modes to the many-body properties of the system. For instance, a common
experimental signature for both pseudogap and precondensation phenomena would be a suppression of entropy
above the critical temperature compared to the high temperature limit. Radio-frequency spectroscopy
allows to deduce the pseudogap regime from the spectral function of cold atomic gases. With this method
a pseudogap regime above the critical temperature has indeed been observed for the UFG \cite{2010NatPh...6..569G,PhysRevLett.106.060402,2014arXiv1409.4743S}.


Moreover, we compare our finding for the precondensation temperature in the
balanced case, $T_{\rm pc}/\mu=0.51$, to analogous values obtained with other
methods. A suppression of the entropy above the critical temperature, and thus a deviation of the
specific heat from the normal gas expectation, is reported in Ref. \cite{PhysRevA.78.023625} 
below $T_0/\mu=0.55(5)$. From the above consideration on the entropy we conclude
that this constitutes an estimate for $T_{\rm pc}$, which is
in good agreement with our result. Our value $T_{\rm pc}/T_{\rm c}=1.25$
is also in line with the results in \cite{PhysRevA.80.033613,PhysRevA.82.043630}
from a T-matrix approach, where different definitions of the pseudogap  have been distinguished.

For vanishing or small spin-imbalance, $\Delta_{0,k}$ approaches zero 
continuously in the precondensation region, see e.g. Fig. 28 in
Ref.~\cite{Boettcher201263}. For configurations with large $\delta\mu/\mu$ and
low $T/\mu$ as in \fig{fig:Deltak}, a jump of $\Delta_{0,k}$ can be observed 
instead. This  behavior is only possible in the vicinity of a first-order phase 
transition. It is generated by a second, non-trivial local minimum 
of the effective potential which is raised above $U_k(\rho=0)$ during the flow 
(cf. insets C and D).
An interesting consequence is that this type of precondensation is not  
necessarily induced by bosonic fluctuations alone. In fact, even in the mean-field
approximation, where the latter are absent, we find a pseudogap regime for large 
$\delta\mu/\mu$, see \fig{fig:phaseMFA}.
 
Furthermore, the peculiar $k$-dependence of the gap at the first-order 
transition region can be exploited numerically. A smooth decrease to zero of 
$\Delta_{0,k}$, as occurring close to a second-order phase transition, may take 
arbitrarily long in RG-time. Therefore, an IR scale of about $t\approx -11$ 
should be considered as an upper limit for the reliable extraction of results 
for finite $\Delta_{0,k=0}$. However, for $T/\mu\leq 0.15$ (shaded area in 
\fig{fig:phaseFRG}), $t\approx -9$ is often the utmost that can be reached, due 
to the increasing stiffness of the flow equations. Thus, the estimate for 
the \emph{value} of $\Delta_{0,k=0}>0$ in the superfluid phase is less 
reliable for such low temperatures. In contrast, the \emph{position} of the 
first-order phase transition is determined by the occurrence of a sudden 
breakdown of the condensate. Indeed, we find that this jump to $\Delta_{0,k}=0$ 
always occurs at some $t> -8$ for $T/\mu \leq 0.15$. Since these scales are not 
affected by the IR problems mentioned above, we conclude that our results for 
the position of the phase transition can be trusted even in the shaded area. 

As a final remark, we mention that the scale evolution of 
$U_k(\rho)$ as shown in \fig{fig:Deltak} allows to check the quality and  
consistency of truncation, regularization and initial conditions. For example, 
it can be seen in inset D that the FRG-evolved effective potential is convex 
for $k\to 0$ within our truncation, cf. \cite{Litim:2006nn}. This exact
property is reproduced by FRG flows~\cite{Ringwald:1989dz,Berges:2000ew}. It
can, however, be spoiled by an insufficient truncation. The mean-field
approximation, for instance, is included  in the FRG equation as a truncation
that neglects all bosonic contributions, cf. our discussion above. However, the
mean-field effective potential is non-convex in the infrared.

\section{Experimental signatures}\label{SecExpSig}
Our findings on the phase structure of the spin-imbalanced UFG have immediate 
consequences on the qualitative shape of in-situ density profiles,
$n(\vec{r})$, obtained for this system in experiment. Here, we briefly
recapitulate the phenomenology of second- and first order phase transitions in
an external potential, and also discuss the impact of the precondensation region
on the interpretation of experimental results. To this end, we define the
density and population imbalance by
\begin{align}
 \label{exp1} 
 n(\mu,\delta\mu,T) &= n_1+n_2 = \Bigl(\frac{\partial P}{\partial 
\mu}\Bigr)_{\delta\mu,T}\,,\\
 \label{exp2} 
  \delta n(\mu,\delta\mu,T) &= n_1-n_2 = \Bigl(\frac{\partial 
P}{\partial \delta\mu}\Bigr)_{\mu,T}\,,
\end{align}
respectively. Here, $P$ is the pressure, and $n_\sigma$ is the density of
atoms in hyperfine state $|\sigma\rangle$\,.

For an ultracold quantum gas confined to an external trapping potential 
$V(\vec{r})$, the thermodynamic equilibrium state depends on the particular
shape of the trap. In many cases, however, we can apply the local density
approximation (LDA), which assigns a local chemical potential
$\mu(\vec{r})=\mu_0 -V(\vec{r})$ to each point in the trap. Here $\mu_0$ is the
central chemical potential. In this way, thermodynamic observables computed for
the homogeneous system are translated into those of the trapped system. Note
that $T$ and $\delta\mu$ are assumed to be constant throughout the trap within
the LDA. The LDA can be applied if the length scale~$\ell_0$ associated to the
trap is much larger than all other scales of the many-body system. For instance,
in a harmonic trap, $V(\vec{r})=M\omega_0^2r^2/2$, the former scale is given by
the oscillator length $\ell_0=\sqrt{\hbar/M\omega_0}$. The many-body length
scales of the UFG are given by $k_{\rm F}^{-1} \propto(2M\mu/\hbar^2)^{-1/2}$,
$(2M\delta\mu/\hbar^2)^{-1/2}$, and $\lambda_T = (M k_{\rm B}T/2\pi
\hbar^2)^{-1/2}$, respectively. Often the LDA is a good approximation for
sufficiently high densities. However, it breaks down in the outer regions of the
trap, where the gas is extremely dilute, and close to a second order phase
transition, where the correlation length diverges (and thus becomes larger than
$\ell_0$).

\begin{figure}
  \includegraphics[width=.4\textwidth]{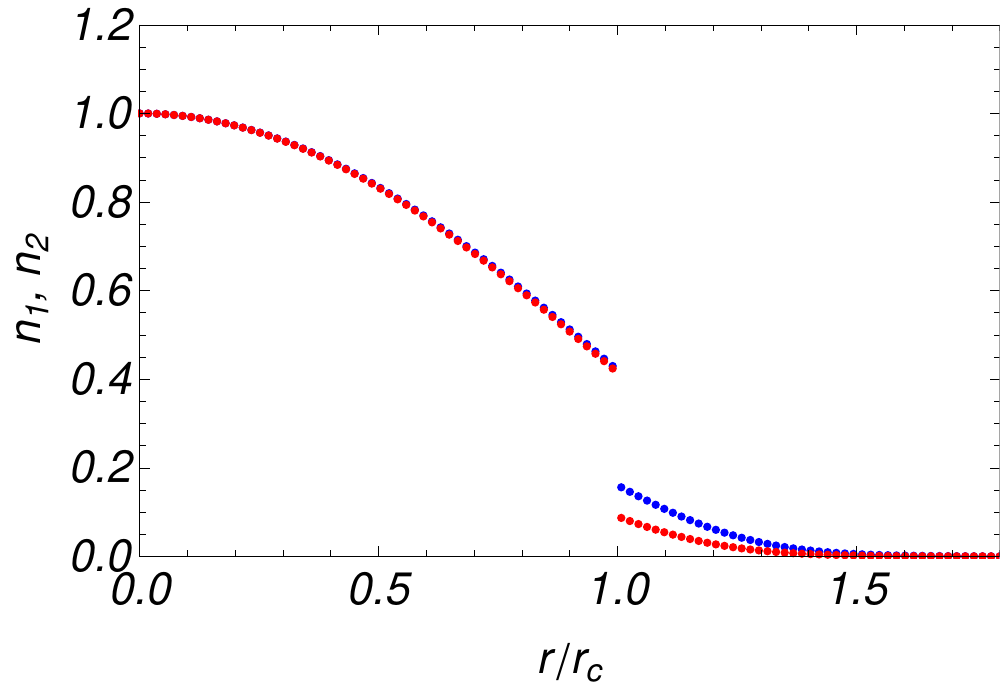}
  \caption{(Color online) Schematic in-situ density profile $n_\sigma(r)$ for a
    population-imbalanced ensemble with $N_1>N_2$ at low temperature. The blue
    and red points correspond to atoms in hyperfine state $|1\rangle$ and
    $|2\rangle$, respectively. For $T<T_{\rm CP}$ the superfluid transition is
    of first order, such that the superfluid inner region is separated from the
    polarized normal gas by a jump in density at the critical radius $r_{\rm
    c}$.}
  \label{FigInsitu}
\end{figure}

If the central chemical potential, $\mu_0$, is sufficiently larger than $T$,
the inner region of the trapped system is superfluid. Above a certain critical
radius, $r_{\rm c}$, the superfluid core vanishes and is replaced by a quantum
gas in the normal phase. The critical radius is related to
the critical chemical potential, $\mu_{\rm c}(T,\delta\mu)$, according to
$\mu_{\rm c}=\mu_0 - V(\vec{r}_{\rm c})$. At a first-order phase transition, the
density at $\mu_{\rm c}$ exhibits a jump. Accordingly, the superfluid inner
region and the normal region are separated by a jump in the density at $r_{\rm
c}$. We sketch this in \fig{FigInsitu}. In contrast, the transition is
continuous for a second-order phase transition. In this way, the order of the
phase transition, and our prediction for the temperature of the tricritical point,
$T_{\rm CP}/\mu_0=0.20$, may potentially be verified from in-situ images at
different temperatures.

In experiments with cold atoms, the imbalance between spin-partners is 
introduced by different atom numbers $N_1 \geq  N_2$ for atoms in state
$|1\rangle$ and $|2\rangle$, respectively. The influence of a non-zero
polarization $p~=~(N_1~-~N_2)~/~(N_1~+~N_2)$ is very distinct for trapped
systems in comparison to homogeneous ones \cite{RLS,Ku:2008vk,Braun:2014ewa}.
For a \emph{trapped system}, the particle numbers $N_\sigma$ are obtained from
an integral over the whole cloud, $N_\sigma = \int_{\vec{r}} n_\sigma(\vec{r})$.
As a consequence, phase separation takes place in real space by means of a
superfluid core and a surrounding normal region. Both are separated by the
above-mentioned jump in the density.

With a state-resolved detection of individual densities, $n_1(\vec{r})$ and 
$n_2(\vec{r})$, it is possible to measure the local in-situ polarization
$p(\vec{r})$ of the trapped gas. According to our finding that there is no Sarma
phase at zero temperature, a non-zero polarization inside the superfluid core of
the cloud can only be detected at $T>0$ \cite{Boettcher:2014xna}. As we find the
Sarma phase only to appear at very high temperatures and close to the phase
boundary, a substantial local polarization $p(\vec{r})$ of the superfluid
should only be detectable for $ r \lesssim r_{\rm c}$. 

To mimic the effect of a trap potential in our RG study, one may consider the
length scale $\ell_0$ associated with a trap as an infrared cutoff, $k_{\rm
f}\sim \ell_0^{-1}$. In fact, in a harmonic oscillator potential, the energy of
the one-particle states is bounded from below by the oscillator
frequency~$\omega_0$. Due to the presence of this infrared scale, long-range
fluctuations are cut off. In a first attempt of simulating trap effects, one may
therefore identify the infrared cutoff $k_{\rm f}$ with $\ell_0^{-1}$ and stop
the RG flow at this scale. However, despite being intuitively reasonable, the
relation $k_{\rm f}=\ell_0^{-1}$ has to be taken with some care and can at best
give qualitative insights~\cite{Braun:2005gy,Braun:2011uq,Braun:2011iz}.

In our analysis we find a substantial precondensation region in the phase 
diagram, \fig{fig:phaseFRG}, where a minimum $\rho_{0,k}>0$ appears during
the flow, but is eventually washed out such that $\rho_0 =\rho_{0,k=0}=0$. The
restoration of symmetry is due to long wavelength fluctuations on length scales
$k^{-1} \to \infty$. However, if  long wavelength fluctuations are cut off by a
trap with scale $\ell_0$, a superfluid order parameter $\rho_0 \approx
\rho_{0,k=k_{\rm f}}$ can be observed experimentally even in the precondensation
phase. 

As discussed in \sec{subsec:ScaleEvol}, the first-order transition is
barely influenced by the final scale $k_{\rm f}$, as long as the latter is below
the relevant many-body scales. Therefore, it may be possible to detect the
first-order transition and its location even in a trap. On the other hand, for
smaller spin-imbalance, where the transition is of second order, this effect can
be substantial. As a consequence, we expect that the second-order phase
boundary of the homogeneous system is likely to be overestimated by applying
the LDA to a trapped gas.

\section{Conclusion and Outlook}
\label{sec:Conclusion}

In this work we have discussed the phase structure of the  spin-imbalanced
unitary Fermi gas as obtained from an FRG study.
This method presents a tool to study the impact of fluctuations in a systematic 
manner. In particular, there is a truncation that is equivalent to the standard 
saddle-point (mean-field) approximation. From this starting point, we have 
additionally included order-parameter fluctuations that are missing in
mean-field theory.

Technically, the discretization of the effective potential on a grid allows us 
to resolve multiple local minima and therefore it opens up the possibility to 
reliably determine first-order phase transitions. Moreover, the full 
functional form of the effective potential is included in such an approach.  

Our results show that the qualitative phase structure persists beyond mean-field
theory: There is a second-order phase transition in the balanced case that 
changes to first order in a tricritical point at finite imbalance. At vanishing 
temperature, superfluidity breaks down in a first-order transition. 
Quantitatively, however, the influence of bosonic fluctuations is more drastic:
in the balanced case, the critical temperature is lowered from $T_{\rm
MF}/\mu=0.665$ to $T_{\rm FRG}/\mu=0.40$, in good agreement with other
theoretical  predictions~\cite{PhysRevA.89.053630}, QMC calculations 
\cite{Wingate:2006wy} and experimental results \cite{Ku03022012}.
At $T=0$ we find that fluctuations enhance the critical imbalance in comparison 
to the mean-field value. This, again, is in line with recent experiments~\cite{PhysRevA.88.063614}. 

Furthermore, the FRG provides access to the full scale-evolution of 
observables, from microscopic to macroscopic scales. This puts us in the 
position to discuss the physics of precondensation, which is related to the 
formation of a condensate at intermediate scales $k$. Interestingly, already
in the mean-field approximation we find a precondensation temperature that is 
significantly higher than the critical temperature at high imbalance. This 
suggests that the formation of a pseudogap is not solely triggered by 
order-parameter fluctuations.
Beyond the mean-field level, the precondensation temperature is substantially 
above the critical one throughout the whole phase diagram. 

Building on the framework presented here, several interesting directions can be 
pursued in the future: For example, it has been conjectured that the UFG 
might feature more exotic phase, such as Sarma~\cite{Sarma19631029,
Gubbels:2006zz, Gubbels:2013mda, Boettcher:2014xna} and/or inhomogeneous (FFLO)
phases \cite{Larkin:1964zz, PhysRev.135.A550, Bulgac:2008tm,
2012arXiv1212.5450B, Gubbels:2013mda,Radzihovsky2012189,Braun:2014}.
Furthermore, the study of mass-imbalance is possible in a similar theoretical 
fashion \cite{Braun:2014} and has gained  experimental interest recently.  

While the UFG is an interesting system that features strong correlations, our
approach is not confined to this setting. The inclusion of finite inverse
scattering lengths is straight-forward and the imbalanced BCS-BEC crossover is
accessible \cite{Boettcher:2014xna}. Moreover, the extension of our approach to 
the two-dimensional BCS-BEC crossover is straight-forward. In the latter case, 
the
importance of a grid-solution for the effective potential is even more
pronounced due to the vanishing canonical dimension of the boson field in two
dimensions.

\begin{center}
 \textbf{Acknowledgements}
\end{center}

\noindent I.B., T.K.H., J.M.P., C.W. thank N. Strodthoff and L. von Smekal for
valuable discussions and collaboration on related projects. I.B. acknowledges
funding from the Graduate Academy Heidelberg. This work is supported by the
Helmholtz Alliance HA216/EMMI and the ERC advanced grant 290623. 
J.B. and D.R. thank J. E. Drut for collaboration on related projects and 
acknowledge support by the DFG under Grant BR 4005/2-1 and by HIC for FAIR
within the LOEWE program of the State of Hesse.

\appendix

\section{Flow equations}
\label{app:Flow}

In this appendix we derive the flow equations for the effective potential and 
the boson anomalous dimensions in the spin-imbalanced UFG. The expressions are 
given in general form (for frequency- and momentum-independent vertices), and
then specialized to our particular choice of truncation and regularization
scheme. In order to simplify the comparison to previous works on the FRG 
approach to the BCS-BEC crossover we remark here that we derive the flow
equations for the unrenormalized couplings only. The latter are often displayed
with an overbar, which we omit here.

The regularized fermion propagator with respect to the field 
$(\psi_1,\psi_2,\psi_1^*,\psi_2^*)$ in our truncation reads
\begin{align}
 \label{flow1} G_\psi(Q) = \frac{1}{\mbox{det}_{F12}^Q\mbox{det}_{F12}^{-Q}} 
\begin{pmatrix} A & B \\ C & D\end{pmatrix}
\end{align}
with
\begin{align}
 \label{flow2} A &= g \phi \begin{pmatrix} 0 & \mbox{det}_{F12}^Q \\ 
    -\mbox{det}_{F12}^{-Q} & 0 \end{pmatrix}, \\
 \label{flow3} B &= \begin{pmatrix} L_{\psi2}^{-Q}\mbox{det}_{F12}^Q & 0 \\
    0 & L_{\psi1}^{-Q} \mbox{det}_{F12}^{-Q} \end{pmatrix}, \\
 \label{flow4} C &= \begin{pmatrix} - L_{\psi2}^Q \mbox{det}_{F12}^{-Q} & 0 \\ 
    0 & - L_{\psi1}^Q \mbox{det}_{F12}^Q \end{pmatrix}, \\ 
 \label{flow5} D &= g \phi \begin{pmatrix} 0 & - \mbox{det}_{F12}^{-Q} \\ 
    \mbox{det}_{F12}^Q & 0 \end{pmatrix}.
\end{align}
We denote $L_{\psi \sigma}^Q = P_\psi(Q)+R_{\psi\sigma}^Q$ and 
$\mbox{det}_{F12}^Q=L_{\psi1}^{-Q}L_{\psi2}^Q+g^2\rho$. For the fermion
regulator we insert either Eq. (\ref{frg7}) or (\ref{frg8}). The regulator
matrix reads
\begin{align}
 \label{flow6} R_\psi(Q) =  \begin{pmatrix} 0 & 0 & - R_{\psi1}^{-Q} & 0 \\
 0 & 0 & 0 & -R_{\psi2}^{-Q} \\ R_{\psi1}^Q & 0 & 0 & 0 \\ 0 & R_{\psi2}^Q & 0 &
0 \end{pmatrix}\,.
\end{align}
The resulting contribution to the flow of the effective potential is given by
\begin{align}
 \label{flow7} \dot{U}^{(F)}(\rho) &= -\frac{1}{2}\mbox{Tr} 
\Bigl(G_\psi\dot{R}_\psi \Bigr)=  - \int_Q
\frac{L_{\psi1}^{-Q}\dot{R}_{\psi2}^{Q} + L_{\psi2}^Q
\dot{R}_{\psi1}^{-Q}}{\mbox{det}_{F12}^Q}.
\end{align}

The regularized boson propagator in the conjugate field basis, $(\phi,\phi^*)$, 
is given by
\begin{align}
 \label{flow8} G_\phi(Q) = \frac{1}{\mbox{det}_B(Q)} 
\begin{pmatrix} -\rho U''(\rho) & L_\phi^{-Q}  \\ L_\phi^Q & -\rho U''(\rho) 
\end{pmatrix}\,,
\end{align}
with $L_\phi^Q =P_\phi(Q)+R_\phi^Q+U'(\rho)+\rho U''(\rho)$ and 
$\mbox{det}_B^Q =L_\phi^{-Q}L_\phi^Q-(\rho U''(\rho))^2$. The corresponding
regulator matrix reads
\begin{align}
 \label{flow9} \bar{R}_\phi(Q) = 
    \begin{pmatrix} 0 & \bar{R}_\phi^{-Q} \\ \bar{R}_\phi^Q & 0 \end{pmatrix}.
\end{align}
We arrive at
\begin{align}
 \label{flow10} \dot{U}^{(B)}(\rho) &= \frac{1}{2}\mbox{Tr} 
\Bigl(G_\phi\dot{R}_\phi \Bigr)= \frac{1}{2}\int_Q
\frac{L_\phi^Q\dot{\bar{R}}_\phi^{-Q}+L_\phi^{-Q}\dot{\bar{R}}
_\phi^Q}{\mbox{det}_B^Q}
\end{align}
for the bosonic contribution to the flow of the effective potential.

We project the flow of the gradient coefficient $A_\phi$ from the 
$\phi_2\phi_2$-component of the inverse boson propagator, i.e. we have
\begin{align}
 \label{flow11} \eta_\phi = -\frac{1}{A_\phi} \frac{\partial^2}{\partial p^2}
\dot{G}^{-1}_{\phi,22}(P)\Bigr|_{P=0,\rho=\rho_{0,k}},
\end{align}
where 
$(\delta^2\Gamma[\phi]/\delta\phi_2\delta\phi_2)(Q,Q')=G_{\phi,22}^{-1}
(Q)\delta(Q-Q')$, 
see e.g. \cite{PhysRevA.89.053630}. In the following  we assume regulators 
which do not depend on the frequency,  $R^Q = R(q^2)$, but the derivation can 
also be performed for $Q$-dependent regulators. We define
\begin{align}
 \label{flow12} R^{\rm x}(q^2) = \frac{\partial R}{\partial q^2}(q^2),\ R^{\rm
xx}(q^2) = \frac{\partial R^{\rm x}}{\partial q^2}(q^2)\,.
\end{align}
We then find $\eta_\phi = \eta_\phi^{(F)}+\eta_\phi^{(B)}$ with
\begin{align}
 \nonumber \eta_\phi^{(F)} &= 2A_\phi g^2 \int_Q \Biggl(  \frac{\dot{R}_{\psi1}(1+
R_{\psi2}^{\rm x}+2q^2R_{\psi2}^{\rm xx}/d)}{(\mbox{det}_{12}^Q)^2}\\
 \label{flow13} &\mbox{ }+\frac{\dot{R}_{\psi2}(1+R_{\psi1}^{\rm x}
    + 2q^2R_{\psi1}^{\rm xx}/d)}{(\mbox{det}_{12}^Q)^2}
    - \frac{4q^2/d}{\mbox{det}_{F12}^Q}\\
 \nonumber & \times\Bigl[\dot{R}_{\psi1}L_{\psi1}^{-Q}(1+R_{\psi2}^{\rm x})^2+
\dot{R}_{\psi2}L_{\psi2}^{Q}(1+R_{\psi1}^{\rm x})^2\Bigr]\Biggr),
\end{align}
and
\begin{align}
 \nonumber \eta_\phi^{(B)} &=4 A_\phi \rho (U'')^2 \int_Q \dot{
\bar{R}}_\phi(q^2) \Biggl(\frac{1+2R_\phi^{\rm x} + 4q^2 R_\phi^{\rm
  xx}/d}{\mbox{det}_B^2(Q)}\\
 \label{flow14} &\mbox{ }-\frac{2q^2(1+2R_\phi^{\rm x})^2
    L_\phi^S(Q)/(A_\phi d)}{\mbox{det}_B^3(Q)}\Biggr)\,.
\end{align}
In the last line we have introduced the symmetric component $L_\phi^S(Q) =
(L_\phi^Q+L_\phi^{-Q})/2$ and employed $d=3$. Note that in order to
evaluate the integrals it is convenient to smear out the step functions
$\theta(x)$ in the regulators, e.g. $\theta_\vare(x)=(e^{-x/\vare}+1)^{-1}$ with
small $\vare>0$. 

For large $k \lesssim \Lambda$ the running of couplings is attracted to an
approximate ultraviolet fixed point where $\eta_\phi=1$
\cite{PhysRevA.76.021602,ANDP:ANDP201010458,PhysRevA.89.053630}. To simplify the
early stage of the flow we start at the fixed point solution. This corresponds
to the initial values
\begin{align}
 g_\Lambda^2 = 6\pi^2 \Lambda,\ \nu_\Lambda = \Lambda^2
\end{align}
within our truncation and regularization scheme. The value for $\nu_\Lambda$ is
fine-tuned such that the resonance condition, $a^{-1}=0$, is fulfilled. The
couplings start to deviate from the ultraviolet fixed point once the flow
parameter reaches the many-body scales, i.e. $k^2 \simeq \mu, T, \delta\mu$. We
choose $\mu/\Lambda^2=10^{-6}$, which is sufficient to suppress the
contributions of many-body effects to the early stages of the flow. The scale
$k^2 =\mu$ then corresponds to an RG-time $t=\ln(\sqrt{\mu}/\Lambda)=-6.9$.

For the optimized cutoffs employed in this work, an overall 
$\dot{R}_{\psi\sigma}$ ($\dot{\bar{R}}_\phi$) implies that
$1+R_{\psi\sigma}^{\rm x}\equiv0$ ($1+2R_{\phi}^{\rm x}\equiv0$) in the
integral. Accordingly, we find
\begin{align}
 \label{flow15} \eta_\phi^{(B)} &=\frac{16 A_\phi \rho (U'')^2}{d} \int_Q 
\dot{\bar{R}}_\phi(q^2)\frac{q^2 R_\phi^{\rm
xx}}{\mbox{det}_B^2(Q)}
\end{align}
for the bosonic contribution to the anomalous dimension. For the fermionic 
contribution the simplification only occurs with the choice
$R_{\psi1}=R_{\psi2}=R_\psi$ with $R_\psi$ from \eq{frg7}. In this case
we arrive at
\begin{align}
 \label{flow16} \eta_\phi^{(F)} &= \frac{8A_\phi g^2}{d} \int_Q \dot{R}_{\psi}  
\frac{q^2R_{\psi}^{\rm xx}}{(\mbox{det}_{12}^Q)^2}\,.
\end{align}

The flow equations for the effective potential and the boson anomalous dimension
can be expressed in closed analytic form for the choice of cutoffs
$R_{\psi1}=R_{\psi2}=R_\psi$ from \eq{frg7} and $R_\phi $ from \eq{frg9}. We
then find
\begin{align}
 \nonumber \dot{U}^{(F)}(\rho) &= -\frac{8 v_d
k^{d+2}}{d\sqrt{1+w_3}}\ell_u(\tilde{\mu}) 
\Bigl(1-N_F(\sqrt{1+w_3}-\delta\tilde{\mu})\\
  \label{flow17} &\mbox{ }-N_F(\sqrt{1+w_3}+\delta\tilde{\mu})\Bigr)
\end{align}
and
\begin{align}
 \nonumber \dot{U}^{(B)}(\rho) &= \frac{4 v_d 2^{d/2} k^{d+2}}{d}
\Bigl(1-\frac{\eta_\phi}{d+2}\Bigr)\frac{2+w_1+w_2}{\sqrt{(1+w_1)(1+w_2)}}\\
 \label{flow18} &\mbox{
}\times\Bigl(\frac{1}{2}+N_B(\sqrt{(1+w_1)(1+w_2)})\Bigr)
\end{align}
with $\delta\tilde{\mu}=\delta\mu/k^2$ and
\begin{align}
 w_1 = \frac{U_k'(\rho)}{A_\phi k^2},\ w_2 = \frac{U_k'(\rho)+2\rho U_k''(\rho)}{A_\phi k^2},\
w_3 = \frac{g^2 \rho}{k^4}\,.
\end{align}
We define $v_d = [2^{d+1}\pi^{d/2}\Gamma(d/2)]^{-1}$, and 
\begin{align}
 N_F(z)=\frac{1}{e^{k^2z/T}+1},\  N_B(z)=\frac{1}{e^{k^2z/T}-1}
\end{align}
with $N_{F/B}'(z)=\partial_z N_{F/B}(z)$, and
\begin{align}
 \ell_u(x) &= \theta(x+1)(x+1)^{d/2}-\theta(x-1)(x-1)^{d/2},\\
 \ell_\eta(x) &= \theta(x+1)(x+1)^{d/2}+\theta(x-1)(x-1)^{d/2}\,.
\end{align}
The contributions to the anomalous dimension read
\begin{align}
 \nonumber \eta_\phi^{(F)} &= \frac{4v_d
A_\phi g^2k^{d-4}}{d(1+w_3)^{3/2}}\ell_\eta(\tilde{\mu})\Biggl[\Bigl(
1-N_F(\sqrt{1+w_3}-\delta\tilde{\mu})\\
 \nonumber &\mbox{ }-N_F(\sqrt{1+w_3}+\delta\tilde{\mu})\Bigr)\\
 \nonumber&\mbox{ }+\sqrt{1+w_3}\Bigl(N_F'(\sqrt{1+w_3}-\delta\tilde{\mu})\\
 \label{flow19} &\mbox{ }+N_F'(\sqrt{1+w_3}+\delta\tilde{\mu})\Bigr)\Biggr]
\end{align}
and
\begin{align}
 \nonumber \eta_\phi^{(B)} &= \frac{8 v_d 2^{d/2} A_\phi \rho_0 (U'')^2
k^{d-4}}{d[(1+w_1)(1+w_2)]^{3/2}}\\
 \label{flow20} &\mbox{ }\times\Bigl(\frac{1}{2}+N_B(\sqrt{(1+w_1)(1+w_2)})\\
 \nonumber &\mbox{ }-\sqrt{(1+w_1)(1+w_2)}N_B'(\sqrt{(1+w_1)(1+w_2)})\Bigr)\,.
\end{align}
In the expressions for $\eta_\phi$ we evaluate the beta functions for
$\rho=\rho_{0,k}$. In the balanced limit, where $\delta\tilde{\mu}=0$, we
recover the flow equations given in Ref.~\cite{ANDP:ANDP201010458}.

\section{Stability of the phase structure}
\label{app:Phase}
\begin{figure}[t!]
  \includegraphics[width=.45\textwidth]{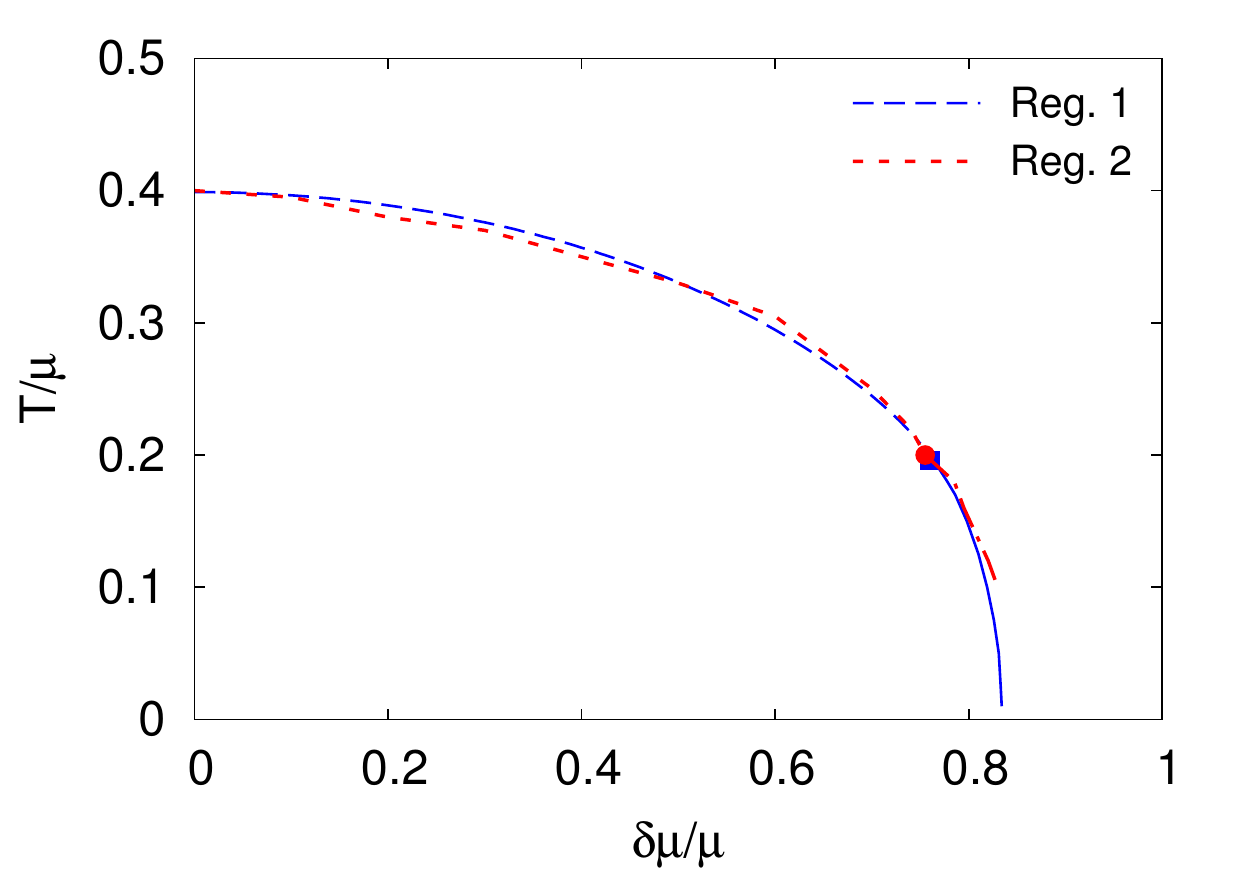}
 \caption{(Color online) Regularization scheme dependence of the phase boundary. We 
  display the phase diagram obtained by applying the fermion regulator
  $R_{\psi\sigma}$ from Eqs. (\ref{frg7}) (blue long-dashed, ``Reg. 1'') and
  (\ref{frg8}) (red short-dashed, ``Reg. 2'').}
  \label{fig:phase_FRG_cmp}
\end{figure}

Here we discuss the stability of the FRG phase structure with respect to
the choice of regulator as well as a different expansion of the flow equation.

As we have discussed in the main text, without truncations to the flow
equations, all permissible regulators should reproduce the
same physics in the IR. In practice, however, one has to resort to truncations 
as well as stop the flow at a finite, if low, infrared scale. This entails that 
the strict regulator-independence is lost. However, for a stable truncation, 
differences should be small. This is what we demonstrate in 
\fig{fig:phase_FRG_cmp}: The blue (solid and dashed)
lines show the superfluid-to-normal transition in the second- and first-order region, 
obtained with the regulator \eq{frg7}. The location of the 
tricritical point is also indicated. The red (short-dashed and dot-dashed) lines 
show the same for regulator \eq{frg8}. As can be seen, the two lines lie close 
to each other throughout the whole phase diagram. Deviations in the critical
temperature are below $5\,\%$ and we mostly attribute them to the presence of
numerical integrals with \eq{frg8}. Hence we can safely claim that our results
are stable with respect to a change in the regulator function.
\begin{figure}[t!]
  \includegraphics[width=.45\textwidth]{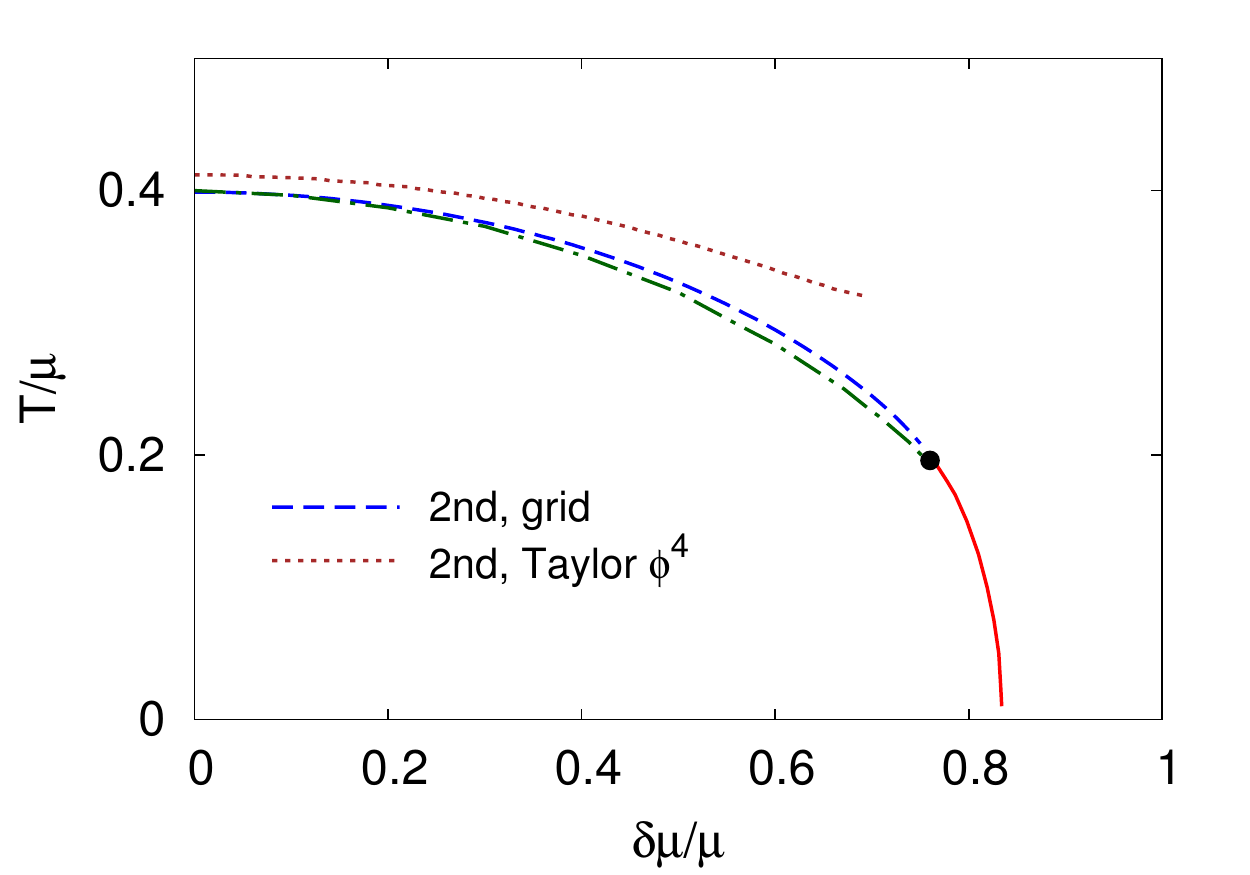}
\caption{(Color online) Using a Taylor expansion of the effective average action $U_k(\rho)$
  to order $\rho^2\sim \phi^4$ (brown, dotted line) in the flow equation, the
  location of the second-order line deviates quantitatively from the grid
  solution (blue, dashed line) as we increase $\delta\mu/\mu$. }
  \label{fig:phaseTaylor}
\end{figure}

Furthermore, we have compared our results to those from a commonly used
truncation scheme for FRG equations: the Taylor expansion of the effective
potential in powers of $\rho-\rho_0$. In \fig{fig:phaseTaylor} we again show our
full phase diagram calculated on a grid, with an added line (brown, dotted)
denoting the result of our Taylor expansion to order $\rho^2\,.$ While the
transition indeed lies close to the grid result at vanishing imbalance (within
$\sim3\,\%$), the deviation increases  to $\sim15-30\,\%$ at high $\delta\mu$\,.
This indicates that, even at low  imbalance, the impact of higher-order terms in
the effective potential is  sizeable.  At high imbalance, the Taylor  expansion
eventually breaks down owing to the presence of an additional minimum  at the
first-order transition.
In this case, the coefficient of $\rho^2$ can turn negative. Since this is the
highest coupling in the system, this entails that the potential becomes
unbounded from below, and hence unstable. Taking into account higher orders in
the Taylor expansion can extend its domain of applicability. To accurately
resolve all minima of the potential, however, very high orders are
needed. Alternatively, one can expand about multiple minima separately or use an
expansion around a fixed value of $\rho$, rather than an expansion around the minimum
\cite{Pawlowski:2014zaa}. Interpreting the breakdown of the Taylor 
expansion as a signal of proximity to the critical point, one would be led to a 
too low $\delta\mu_{\rm CP}$ as well as too high $T_{\rm CP}$\,, at least in 
this low expansion order. An expansion of the effective average potential to
order $\phi^4$ has recently been applied to a spin-imbalanced Fermi gas with
weak attractive interactions \cite{Krippa:2014kra}. Accordingly, the superfluid
transition was found to be of second order.
While the inclusion of higher order terms might diminish the discrepancy in the
second-order line to some extent \cite{Pawlowski:2014zaa}, the  resolution of
the first-order transition is more challenging within a Taylor expansion. 


\bibliographystyle{apsrev4-1}
\bibliography{references_imbalance} 

\end{document}